

\input phyzzx.tex


\def\np{Nucl. Phys.}
\def\pl{Phys. Lett.}

\def\cmp{Commun. Math. Phys.}
\def\intmp{Intern. J. Mod. Phys.}

\def\dd{\hbox{d}}


\Pubnum={SISSA/77/93/EP}
\date{June 1993}

\titlepage

\title{\bf The BRST quantization of the nonlinear $WB_2$ and $W_4$ algebras}
\author{Chuan-Jie Zhu}
\address{ International School for Advanced Studies (SISSA/ISAS)
\break {\rm and} \break INFN, Sezione di Trieste \break
via Beirut 2-4, I-34013 Trieste, Italy }

\vglue .5cm

\abstract\nobreak
{We construct the BRST operator for the non-linear $WB_2$ and $W_4$ algebras.
Contrary to the general belief, the nilpotent condition of the BRST operator
doesn't determine all the coefficients. We find a three and seven parameter
family of nilpotent BRST operator for $WB_2$ and  $W_4$ respectively. These
free parameters are related to the canonical transformation of the ghost
antighost fields. }

\vfill\eject
\endpage

\centerline{ \bf 1. Introduction }

\vglue .5cm

\REF\WIT{ For review see for example:
M. B. Green, J. H. Schwarz and E. Witten, {Superstring theory},
2 vols. (Cambridge University Press, 1987) }
It is well-known that the covariant quantization of the (bosonic) string
theory leads to  a set of constraints for the physical state [\WIT]
$| \hbox{ phys.} \rangle $:
$$
L_n | \hbox{ phys.} \rangle = 0, \qquad\qquad  n \geq 0 , \eqn\onea
$$
where $L_n$'s are the modes of the stress-energy tensor
$T(z) = \sum_n L_n z^{-n-2} $ and satisfy the following Virasoro
algebra
$$
[ L_m, L_n ] = (m-n) L_{m+n} + { D \over 12} m (m^2 - 1)
\delta_{m+n,0} , \eqn\oneb
$$
or in terms of $T(z)$, the following OPE:
$$
T(z) T(w) \sim { D\over 2(z-w)^4 } + \left(
{ 2\over (z-w)^2}  + { 1\over z-w} \partial_w \right) T(w).
\eqn\onec
$$
\REF\BRSA{C. Becchi, A. Rouet and R. Stora, Ann. Phys. {\bf 98}
(1976) 287; \cmp\ {\bf 42} (1975) 127 }
\REF\BRSB{I. V. Tyutin, Gauge invariance in field theory and statistical
mechanics, Lebedev preprint FIAN No. 39 (1975) unpublished }
\REF\KAT{M. Kato and K. Ogawa, \np\ {\bf B212} (1983) 443 }
One way of organizing all these constraints is to introduce the
ghost modes $c_m$ and their corresponding antighost modes $b_n$
which satisfy the following anti-commutation relations
$$
\eqalign{
\{ c_m, b_n \} = &  \delta_{m+n, 0} , \cr
\{ c_m, c_n \} = & \{ b_m, b_n \} = 0 .
} \eqn\oned
$$
Then by constructing a BRST operator (firstly introduced in the
quantization of gauge fields [\BRSA, \BRSB]) [\KAT]
$$
Q = \sum_n c_{-n} L_n - { 1\over 2 } \sum_{m,n} : c_{-m}c_{-n}b_{m+n}: ,
\eqn\onee
$$
the physical state conditions can be formulated as the following
$$
Q | \hbox{ phys.} \rangle = 0  \qquad \hbox{ and } \qquad
| \hbox{ phys.} \rangle \neq Q| \chi  \rangle .
\eqn\onef
$$
\REF\GRA{B. Lian and G. Zuckermann, \pl\ {\bf B254} (1991) 417;
\pl\ {\bf B266} (1991) 21; \cmp\ {\bf 561} (1992) 561 }
\REF\GRB{P. Bouwknegt, J. McCarthy and K. Pilch, \cmp\
{\bf 145} (1992) 541 }
Here the BRST operator $Q$ is required to be nilpotent, i.e. $Q^2 = 0 $,
because of the consistency of the formalism. This requirement fixes
$ D = 26 $, the celebrated critical dimension of the bosonic
string theory  [\KAT, \WIT].
If we identify $Q$ as an exterior differential operator, from $\onef$
we see that the physical states are just these closed but not exact
differentials, i.e. the cohomology classes. The payoff of using the BRST
formalism is that one can use the well-established results in cohomology
theory to compute the physical states. In fact a complete understanding
of the physical states in minimal model coupled with gravity was
reached by computing the cohomology of the corresponding BRST operator
[\GRA, \GRB].

\REF\KZ{V. G. Knizhnik and A. B. Zamolodchikov, \np\ {\bf B247} (1984) 83 }
\REF\ZAM{A. B. Zamolodchikov, Theor. Math. Phys. {\bf 65} (1985) 1205}
\REF\MIN{J. Thierry-Mieg, \pl\ {\bf B197} (1987) 368 }
\REF\BRSTA{M. Bershadsky, W. Lerche, D. Nemeschansky and N. P. Warner,
\pl\ {\bf B292} (1992) 35}
\REF\BRSTB{E. Bergshoeff, A. Sevrin and X. Shen, \pl\ {\bf B296} (1992) 95}
\REF\MINA{K. Schoutens, A. Sevrin and P. van Nieuwenhuizen,
\cmp\ {\bf 124} (1989) 87}
\REF\POPE{H. Lu, C. N. Pope and X. J. Wang, On higher-spin generalizations
of string theory, preprint CTP-TAMU-22/93 (April 1993) }
\REF\HORN{K. Hornfeck, Explicit construction of the BRST charge for $W_4$,
preprint DFTT-25/93 (May 1993) }

The BRST formalism is quite powerful and has been extended to many
algebras. Examples are superconformal algebras which appear in
superstring theory [\WIT]
and current algebra which appears in Wess-Zumino-Witten
model [\KZ].
For all these algebras a (quantum) BRST operator can often easily be
constructed. Up to some anomalous terms the Jacobi identities of the
algebras guarantee the nilpotence of $Q$. (The vanishing of the
anomalous terms gives the critical dimension but this  is not always
possible.) This is so because all these algebras are (graded) linear
algebras. On the other hand, for non-linearly extended algebras
the construction of  quantum BRST operators is a much more difficult
problem. For the simplest non-linear algebra $W_3$ [\ZAM], the BRST
operator has been constructed in [\MIN] and shown to be nilpotent
if the central charge is 100. (See also refs. [\BRSTA, \BRSTB].)
By understanding this construction, in
[\MINA] a general solution was given for the construction of
quantum BRST operator for a sub-class of quadratical non-linear algebras.
Nevertheless such restriction on the algebra is so stringent that even
the next simplest non-linear $W$-algebras $WB_2$ and $W_4$ are not belong
to the class\footnote*{These two non-linear algebras contain tri-linear term
$\Lambda_7$ ($\sim T^3$) in the conformal basis, but they are actually
quadratic non-linear algebras.}. The BRST operator for the $WB_2$ algebra was
studied in ref. [\POPE] by using an explicit free field realization for
part of the stress energy tensor.  In this paper we will construct the
(quantum) BRST operator  for these two simple non-linear algebras from a
purely algebraic point of view. The motivation lies behind this study
is that we hope to learn something new from these simple cases.
Our results hint that no general recipe could exist for the
construction of BRST operator for these  non-linear algebras.
Contrary to the general belief we found that there
exists not just a unique BRST operator but a family of BRST operators.
In the case of $WB_2$ we found a three parameter family of nilpotent
BRST operator. For $W_4$ there is a seven parameter family of nilpotent
BRST operator. Independently the BRST operator for the $W_4$ algebra was also
constructed in ref. [\HORN]. It was pointed out in ref. [\HORN] that
these free parameters are related to the canonical transformation of the
ghost antighost fields. We will establish this connection explicitly in
sect. 5 for the $WB_2$ algebra.

\REF\ZHU{C. J. Zhu, in preparation }
\REF\INDIA{S. Mukherji, S. Mukhi and A. Sen, \pl\ {\bf B266} (1991) 337}
\REF\ASK{S. R. Das, A. Dhar and S. Kalyana Roma, \intmp\ {\bf A7}
(1992) 2295}
\REF\WWA{R. Blumenhagen, M. Flohr, A. Kliem, W. Nahm,
A. Recknagel and R.  Varnhagen, \np\ {\bf B361 } (1991) 255 }
\REF\WWB{H. G. Kausch and G. M. T. Watts, \np\ {\bf B354} (1991) 740 }
\REF\ZHUOVER{C. J. Zhu, The complete structure of the nonlinear $W_4$ and
$W_5$ algebras from quantum Miura transformation,
SISSA preprint SISSA/76/93/EP (June 1993) }

Before  presenting the
results let me make a few remarks about this work and the writing of this
paper. This may be helpful to understand this paper.
Based on my study about the cohomology of pure gravity and subsequently
some preliminary works on pure $W_3$ gravity [\ZHU], I envisage that
there should be a general connection between highest states and BRST
cohomology in pure $W$-gravity [\INDIA, \ASK].
Central to this conjecture is the
assumption that there exists a (unique quantum) BRST operator.
So I set to construct the BRST operator for the $W_4$ algebra.
In order to fix the notation I rederived the $W_4$ algebra [\WWA, \WWB] from
quantum Miura transformation [\ZHUOVER]. The construction of the BRST operator
is straightforward although the algebraic calculations are so complicated
that the only hope is to use computer symbolic calculation. With the aid of
computer, all the coefficients are found and the BRST operator is shown to be
nilpotent. Nevertheless the result obtained is  quite dirty: the printout
(in phyzzx TeX form) of the coefficients is around 20 pages long and
a seven parameter family of nilpotent BRST exists\footnote*{In an early
version of this paper, I got only a two parameter family by requiring some
total derivatives to zero. The solution found is only a subset of the
complete solution.}. Next I realized that
there is a $WB_2$ algebra which is simpler than $W_4$ algebra because
there is no spin-3 fields. With $WB_2$ algebra life becomes much easier.
Nevertheless there still exists a three parameter family of nilpotent
BRST operator. As we will show in sect. 5, by requiring the total
stress energy tensor as $\{Q, b(z)\}$ we can fix some of these free
parameters and the expression of $Q$ simplifies a lot. For $WB_2$, all the
three free parameters are fixed  and we have a unique BRST operator. For
$W_4$ there left only two free parameters and there seems no natural choice
for them. The organization of the paper is as follows:

In sect. 2 we gave all the needed OPEs for the $WB_2$ and $W_4$ algebras.
The OPEs involving the composite field $\Lambda_1$ are also given
explicitly. In sect. 3 we discuss how one can construct a BRST operator
for any given algebra. Specifically we gave the ansatz of the BRST currents
for the $WB_2$ and $W_4$ algebras. In sect. 4 the solutions of the
nilpotent condition $Q^2= 0 $ are given. Nevertheless we gave
all the coefficients only for $WB_2$ and dare not to write down
all the coefficients for $W_4$. The best we could do is to fix five out
of the seven free parameters and gave only the resulting expressions
which depend on only two free parameters. In sect. 5 we discuss the results
obtained in sect. 4 and explicitly establish the connection between these
free parameters and the canonical transformation of the ghost antighost
fields which was only briefly touched in ref. [\HORN]. The explcit form of
all these canonical transformations are also given.

\vfill\eject

\centerline{ \bf 2. The nonlinear $WB_2$ and $W_4$ algebras }

\vglue .5cm
\REF\WBA{P. Bouwknegt, \pl\ {\bf B207} (1988) 295 }
\REF\WBB{K. Hamada and M. Takao, \pl\ {\bf B209} (1988) 247 }
\REF\WBC{D. H. Zhang, \pl\ {\bf B232} (1989) 323 }

In this section we gave the complete OPEs of the nonlinear $WB_2$ and $W_4$
algebras. We will use the same notation for both algebras (but different
normalization for the spin-4 field $U(z)$). Firstly the
$WB_2$ algebra is
generated by the stress-energy tensor $T(z)$ and a spin-4 primary field
$U(w)$. The basic OPEs were discussed in a number of papers
[\WBA-\WBC, \WWA, \WWB]. They are
$$
\eqalign{
T(z) T(w)   \sim &  \Big( { 2 \over (z-w)^2 } + { 1\over z-w} \partial_w
+{ 3\over 10  } \partial_w^2 + { 1\over 15 } (z-w) \partial_w^3
 + { 1\over 84 } (z-w)^2 \partial_w^4 \Big) T(w) \cr
&  + { c/2 \over (z-w)^4 } +
\Big( 1 + { 1\over 2} (z-w) \partial_w + { 5 \over 36 } (z-w)^2 \partial_w^2
\Big) \Lambda_1(w) + (z-w)^2 \Lambda_2(w) , \cr
& \cr
T(z) U(w) \sim &   \Big( { 4\over (z-w)^2 } + { 1\over z-w} \partial_w +
{ 1\over 6 } \partial_w^2 \Big) U(w) + \Lambda_5(w) ,  \cr
& \cr
U(z)U(w) \sim & { c/4\over  (z-w)^8 } +
\Big( { 2\over (z-w)^6 }  + { 1\over (z-w)^5 } \partial_w +
{ 3/10 \over (z-w)^4 } \partial_w^2 + { 1/15 \over (z-w)^3 } \partial_w^3 \cr
& + { 1/84 \over (z-w)^2 } \partial_w^4 \Big) T(w)
+ \Big(  { 1\over (z-w)^4 } + { 1/2 \over (z-w)^3 } \partial_w
+ { 5 /36 \over (z-w)^2 } \partial_w^2  \Big) \cr
& \times \Big(  c_0 U(w) + { 42 \over (22 + 5c)}  \Lambda_1(w) \Big) +
{ 28c_0\over 3(c + 24) } { \Lambda_5(w) \over (z-w)^2 } \cr
& + { 3(19c - 524) \over 5 (68 + 7c)(2c-1) } { \Lambda_2(w) \over
(z-w)^2 } + { 24 (72c + 13) \over (22 + 5c) (68 + 7c) (2c - 1) }
{ \Lambda_7(w) \over (z-w)^2 }  , } \eqn\opewb
$$
where $c_0 = \sqrt{ 54 (24 + c) ( c^2 - 172 c
+ 196 ) \over (22 + 5c) (68 + 7c)
(2c- 1) }$ . Notice that in $\opewb$ we didn't give the simple pole terms
in $U(z)U(w)$ because these terms can easily be
obtained from the symmetric property of this OPE
and they are not needed explicitly in the following. Also some regular terms
are included in $T(z)T(w)$ and $T(z)U(w)$ explicitly in order
to define the quasi primary
fields $\Lambda_i(w)$ ($i$ = 1, 2, 5). The other quasi primary field
$\Lambda_7(w)$ appears in the regular term of the following OPE:
$$
T(z) \Lambda_1(w) \sim  { (22 + 5c) \over 5 } { T(w) \over (z-w)^4 }
+ \Big( { 4\over (z-w)^2 } + { 1\over z-w } \partial_w
+ { 1\over 6 } \partial_w^2 \Big) \Lambda_1(w) + \Lambda_7(w) .
\eqn\opea
$$
\REF\REVI{For review see for example:
P. Bouwknegt and K. Schoutens, Phys. Reps. {\bf 223} (1993) 183}
In the computation of $Q^2$ we will also need the OPEs between
$U(z)$ and $\Lambda_1(w)$, and $\Lambda_1(z)$ with itself because
the composite field $\Lambda_1(z)$ also enters  the construction
of $Q$ due to the presence of a tri-linear term $\Lambda_7(w)$ in
$U(z)U(w)$. These
OPEs can be computed from the Wick theorem involving the contraction
of composite fields [\REVI]. Explicitly we have
$$\eqalign{
U(z) \Lambda_1(w) \sim &
{ 84 \over 5 }  \Big({ 1 \over (z-w)^4 } + { 1/2 \over (z-w)^3 }
\partial_w + { 5/36 \over (z-w)^2 } \partial_w^2
\Big) U(w) + { 8 \over (z-w)^2 } \Lambda_3(w) , \cr
& \cr
\Lambda_1(z) \Lambda_1(w) \sim & { c(22+5c)\over 10 (z-w)^8 }
+ { 4 (22+5c) \over 5 } \Big( { 1\over (z-w)^6 } +
{ 1/2 \over (z-w)^5 } \partial_w \cr
& + { 3/20 \over (z-w)^4 } \partial_w^2 +
{ 1/15\over (z-w)^3 } \partial_w^3+{ 1/168\over (z-w)^2 } \partial_w^4\Big)T(w)
\cr & +  { 2(64 + 5c)\over 5 } \Big(
{ 1\over (z-w)^4} + { 1/2 \over (z-w)^3} \partial_w
+ { 5/36 \over (z-w)^2} \partial_w^2 \Big) \Lambda_1(w) \cr
& + { 2 (22 + 5c) \over 5 } { \Lambda_2(w) \over (z-w)^2 }
  + { 8 \over (z-w)^2 } \Lambda_7(w) . }
 \eqn\opeb $$

For $W_4$ algebra there is a spin-3 primary field $W(z)$  besides the
stress-energy  tensor $T(z)$ and the spin-4 primary field $U(z)$ appearing
in $WB_2$ algebra.
However the $WB_2$ algebra is not a sub-algebra of the $W_4$ algebra.
The OPE $U(z)U(w)$ in $W_4$ algebra is different from the one in $WB_2$
algebra because the presence of an additional spin-6 composite field
$\Lambda_6 \sim W^2$.
The complete structure of the $W_4$ algebra was determined by using Jacobi
identities in [\WWA, \WWB]. Recently we have rederived it from quantum Miura
transformation [\ZHUOVER]. For later convenience we will write the OPEs in a
non-standard (but natural from quantum Miura transformation) normalization.
The OPEs $T(z)T(w)$, $T(z)U(w)$, $T(z)\Lambda_1(w)$, $U(z)\Lambda_1(w)$ and
$\Lambda_1(z)\Lambda_1(w)$ are the same as those given above for $WB_2$. The
OPE $U(z)U(w)$ changed to the following
$$ \eqalign{
{U(z)U(w) \over C_4 }  \sim &
 { c/4 \over  (z-w)^8 } +
\Big( { 2\over (z-w)^6 }  + { 1\over (z-w)^5 } \partial_w +
{ 3/10 \over (z-w)^4 } \partial_w^2 + { 1/15 \over (z-w)^3 } \partial_w^3 \cr
& + { 1/84 \over (z-w)^2 } \partial_w^4 \Big) T(w)
+ \Big( { 1\over (z-w)^4 } + {1/2\over (z-w)^3 } \partial_w
+ {5/36 \over (z-w)^2 } \partial_w^2 \Big) \cr
& \times \Big( { 42 \over (22 + 5c) } \Lambda_1(w)
- { 90 (c^2 + c + 218 ) \over (2 +c) (7+c) (7c +114) }  U(w) \Big) \cr
& - { 120\over (2+c)(7+c) } { \Lambda_5(w)\over (z-w)^2 }
 + { 225 (22 + 5c) \over (2 + c)(7+c)(7c + 114) }
{ \Lambda_6(w) \over (z-w)^2} \cr
& + { 3 (19c - 582)\over 10 (2+c)(7c + 114)}
{ \Lambda_2(w) \over (z-w)^2 }
+ { 96 (9c - 2) \over (2 +c)(22 + 5c)(7c +114) }
 { \Lambda_7(w) \over (z-w)^2}
, } \eqn\opeuu $$
where  $ C_4 = { (2 +c)(7+c)(7c + 114)\over 300 (22 +5c) }  $.
The additional OPEs $T(z)W(w)$, $W(z)W(w)$, $W(z)U(w)$ and $W(z)\Lambda_1(w)$
are given as in the following:
$$ \eqalign{ T(z) W(w) \sim & ~~
\left( { 3\over (z-w)^2 } + { 1\over z-w} \partial_w +
{ 3\over 14} \partial_w^2 + { 3\over 84} (z-w)\partial_w^3 \right) W(w) \cr
& + \big( 1 + { 2\over 5} ( z-w) \partial_w\big) \Lambda_3(w)
+ (z-w) \Lambda_4(w), \cr
& \cr
{ W(z) W(w) \over ( 7 +c)/10 } \sim & ~~ { c/3 \over (z-w)^4 }
+ \Big( { 2\over (z-w)^4 }  + { 1\over (z-w)^3 } \partial_w +
{ 3/10 \over (z-w)^2 } \partial_w^2 + { 1/15 \over z-w } \partial_w^3 \cr
& + { 1\over 84 } \partial_w^4 \Big) T(w)
+ { 32 \over (22 + 5c) }
\Big( { 1\over (z-w)^2 } + {1/2\over z-w} \partial_w
+ {5\over 36} \partial_w^2 \Big) \Lambda_1(w) \cr
& + {40\over (7+c) }
\Big( { 1\over (z-w)^2 } + {1/2\over z-w} \partial_w
+ {5\over 36} \partial_w^2 \Big) U(w) + { 10\over (7+c) } \Lambda_6(w) , \cr
& \cr
W(z) U(w) \sim &   { (c+2)(7c + 114) \over 10 (22+5c) }
\Big( {1\over (z-w)^4 }  + {1/3 \over (z-w)^3 } \partial_w
+ { 1/14 \over (z-w)^2 } \partial_w^2
+ { 1/84 \over z-w} \partial_w^3 \Big) W(w) \cr
& + { 26 ( c+ 2) \over 5 (22 + 5c)}
\Big( { 1 \over (z-w)^2 } + { 2/5 \over z-w } \partial_w \Big) \Lambda_3(w)
+  { (7c + 114) \over 10 (22 + 5c)} { \Lambda_4(w) \over (z-w) } , \cr
& \cr
W(z)\Lambda_1(w) \sim & ~~{ 48 \over 5 } \left( { 1\over (z-w)^4} +
{ 1/3 \over (z-w)^3 } \partial_w + { 1/14 \over (z-w)^2 } \partial_w^2 +
{ 1/84 \over z-w} \partial_w^3 \right) W(w) \cr
& + 6 \left( { 1\over (z-w)^2 } + { 2/5 \over z-w} \partial_w \right)
\Lambda_3(w)  - { 4\over (z-w) } \Lambda_4(w) . } \eqn\opewl $$

\vfill\eject

\centerline{ \bf 3. The Ansatz }

\vglue .5cm

Following the standard procedure we introduce ghost anti-ghost
pairs $(c(z), b(z))$, $(\gamma(z)$, $\beta(z))$
and  $(\delta(z), \alpha(z))$ for $T(z)$, $W(z)$  and $U(z)$
respectively. These ghost anti-ghost fields have spins $(-1,2)$,
$(-2,3)$  and
$(-3,4)$ and their mode expansions are as follows
$$\eqalign{
c(z) = & \sum_n c_n z^{ -n + 1} , \cr
\gamma(z) = & \sum_n \gamma_n z^{-n+2}, \cr
\delta(z) = & \sum_n \delta_n z^{ - n  + 3} , \cr}
\qquad\qquad
\eqalign{
b(z) = & \sum_n b_n z^{ -n - 2} , \cr
\beta(z) = & \sum_n \beta_n z^{-n-3}, \cr
\alpha(z) = & \sum_n \alpha_n z^{ - n  - 4} . } \eqn\modea $$
These modes satisfy the usual anti-commutation relations which can be derived
from the following OPEs
$$\eqalign{
c(z) b(w) \sim &   { 1\over z-w}, \cr
\gamma(z)\beta(w) \sim & { 1\over z-w} , \cr
\delta(z) \alpha(w) \sim & { 1\over z-w} . }   \eqn\modeb $$
The other OPEs are all 0.

Because of the complexity with normal ordering we will not use
mode expansions. All our calculation are done with (the holomorphic)
fields. The normal ordering for the ghost anti-ghost fields are such
that the following equations are true
$$
\eqalign{
c(z) b(w) =&  { 1 \over z-w } + : c(z) b(w) : , \cr
\gamma(z)\beta(w) = & { 1\over z-w} + : \gamma(z) \beta(w) :, \cr
\delta(z) \alpha(w) =& { 1\over z-w } + : \delta(z) \alpha(w) : . }
\eqn\modec$$
This is possible because all these fields are free fields.

\REF\BCC{ M. Henneaux, Phys. Rep. {\bf 126} (1985) 1 }
\REF\BCD{M. Henneaux and C. Teitlboim, Quantization of gauge systems,
(Princeton University Press, 1992) }

With all the above knowledge, we now construct the quantum BRST
operator. One way to start is to construct the corresponding classical
BRST operator [\BCC, \BCD].
The quantum BRST operator is then assumed to be the same
form as the classical one with possible renormalization of some
coefficients and addition of some zero mode terms due to normal
ordering. By imposing the nilpotence condition, one could determine
all these coefficients. For linear algebras this route is quite successful.
The same strategy has  been applied to $W_3$ [\MIN]
and in [\MINA] to a class of quadratic non-linear algebra. But the
simplicity of this construction doesn't apply to  $WB_2$ and other non-linear
algebras. As we will show explicitly a moment later, there are new terms which
are not present in the classical BRST operator. So we should look for other
method.

{}From our experience with BRST operator in known examples we know
that the (quantum) BRST operator is the contour integration of a spin-1
current $j(z)$ with ghost number 1. $j(z)$ is the summation of various terms.
Each individual term is the (normal ordered) product of the basic fields and
their derivatives. For example, the BRST operator in
$\onee$ is the integration of $j(z) = c(z) T(z) + c(z) \partial_z c(z)
b(z) $: $Q = \oint_0 [\dd z] j(z) $\footnote*{ $[ \dd z ] \equiv { \dd z  /
2 \pi i } . $ }. ($j(z)$ is called the BRST current.)
We can simply assume that this is
true in general. As the BRST operator is presumably coming from
the quantization of a constrained system, we will also assume that
the single ghost terms in $j(z)$ is given by $\sum_i c_i(z) T_i(z)$,
where $T_i(z)$'s generate the constraints and $c_i(z)$'s are the corresponding
ghost fields. For $WB_2$ algebra this is $c(z)T(z) + \delta(z) U(z) $.
There is no terms like $\delta(z) \partial_z^2 T(z)$ which also has
spin 1. (However see discussion in sect. 5.)  Then we can just
write down the most general spin-1 current with ghost number 1 (this is
possible because these terms are finite in number, see below). Presumably the
nilpotence of the BRST operator constructed from this current by contour
integration should fix all the unknown coefficients.
Let us now apply these rules to the construction of BRST currents for
the $WB_2$ and $W_4$ algebras.

First let us see how we can generate all the possible terms for $WB_2$.
The ghost
number condition put the constraint that an individual term should consist
of $n+1$ ghost fields and $n$ anti-ghost fields with $n=1$, 2, $\cdots$.
Define a generating function
$$
P(x) =  \sum_{ n= 1}^{\infty} : \Big( { c(w+x) \over x} + { \delta(w + x)
\over x^3 } \Big)^{n+1} (x^2 b(w+x) + x^4 \alpha (w+x) )^n :
 \equiv  \sum_l P_l x^l .  \eqn\klk
$$
Here we should first think these ghost anti-ghost fields as commutating.
Only at the end of the expansion we consider them as anti-commutating
and put $(c(w))^2$, $(\partial_w b(w))^3$, etc. to 0.
It is not quite difficult to convince oneself that $P_l$ is a spin-$l$
current with ghost number 1. Because all
the bosonic fields have positive spins, only the  currents $P_l$
with $l\leq 1$ could possibly be included in BRST current $j(z)$. Due to the
anti-commutativity of the ghost fields, these currents are finite in number.
For $l < 1$, one can construct spin-1 fields from $P_l$ by multiplying with
bosonic
fields. We claim that these are all the possible terms. Of course every $P_l$
consists of several terms and their coefficients should be set free in $j(z)$.

After describing the general principle let us see what is the maximum number
of ghost fields a term can have. Because all the non-positive spin ghost
fields are $(\delta(z),$  $\delta'(z),$ $\delta''(z),$ $c(z),$
$\delta^{(3)}(z),$
$c'(z))$, the lowest spin (first consideration) fields with lowest ghost
number (second consideration) is $\delta(z) \delta'(z) \delta''(z)c(z)$
and  has
spin $- 7 $ and ghost number 4. Nevertheless the lowest spin field with
ghost number $-3$ are $b(z)b'(z)b''(z)$ and $b(z)b'(z)\alpha(z)$ and have
spin 9.
So there is no way to construct a spin-1  field by taking the product.
The next possibility is to have 3 ghost fields and 2 anti-ghost field. Here
there are a lot of terms. Some of them are
$$\eqalign{
& b(z)b'(z)\delta(z)\delta'(z)\delta''(z)T(z), \cr
& c(z)c'(z)b(z)b'(z)\delta(z), \cr }
\qquad\quad
\eqalign{
& c(z) b(z)b'(z)\delta(z)\delta'(z)T(z) , \cr
& b(z)b'(z)\delta(z)\delta'(z)\delta^{(4)} . \cr }
\eqn\ong
$$

One subtle point in the above construction of the BRST current is that there is
a redundancy of terms because the integration of a total derivative term
identically gives zero. So we should set some terms to zero in order to fix
this
redundancy. Bearing this in mind, the most general expression of the BRST
current $j(z)$ for $WB_2$ can be easily constructed. Splitting it as the sum
of various term of fixing total number of ghost antighost fields, we have
$$ j(z) = j_0(z) + j_1(z) + j_2(z) , \eqn\oooa $$
where the single ghost term is
$$
j_0(z) =  c(z) T(z) + \delta(z) U(z) . \eqn\sina$$
The three ghost antighost term
$j_1$ is\footnote*{ All the fields $b$, $c$, $\delta$,
$\alpha$ and their derivatives are holomorphic function of $z$. }
$$\eqalign{
j_1 = &  ~~  b\delta\delta' (a_0 \Lambda_1 + a_1 U)
+ \big( a_2 b\delta\delta^{(3)} + a_3 b \delta' \delta''
+  a_4 b' \delta\delta''  + a_5 b'' \delta \delta' + a_6cb\delta'
+ a_7 cb'\delta \cr
& + a_8 c' b \delta + a_9 \delta\delta' \alpha \big) T
+ \big( m_1\delta\delta^{(5)}  + m_2 \delta' \delta^{(4)}
+ m_3 \delta'' \delta^{(3)}  \big) b
+ c \big( m_4 \delta' \alpha + m_5 \delta\alpha' \big)   \cr
& +  m_6 cc'b  + c \big( m_7 b'\delta'' + m_8 b \delta^{(3)}
+ m_9 b'' \delta' + m_{10} b^{(3)} \delta \big)
+ \delta \big( m_{11} \delta' \alpha'' + m_{12}\delta^{(3)}
\alpha \big), } \eqn\pcpx$$
and the five ghost antighost term $j_2$ is
$$\eqalign{
j_2 = & ~~\big( a_{10} b b'\delta \delta' \delta''
 + a_{11} cb b' \delta\delta' \big)  T + m_{13}c c' b b' \delta
+ b\delta\big( m_{14} \delta' \delta'' \alpha'
+ m_{15} \delta'\delta^{(3)} \alpha \big) \cr
& + b\big( m_{16} b' \delta' \delta^{(4)} + m_{17} b'
\delta''\delta^{(3)} + m_{18} b^{(3)}  \delta'\delta''
\big)\delta  + c\big( m_{19} b' \delta' \alpha + m_{20}b\delta'\alpha'
+ m_{21} b \delta''\alpha \big) \delta \cr
& + c \big( m_{22} bb' \delta \delta^{(3)} +
m_{23} bb' \delta'\delta''  + m_{24} bb'' \delta\delta''
+ m_{25} bb^{(3)}  \delta \delta' + m_{26} b'b'' \delta\delta' \big) .  }
\eqn\jjz $$
There is no 7 or higher ghost antighost terms.
Notice that in $j_1(z)$ there also appears a term with $\Lambda_1(z)$ (
$\sim T^2(z)$). This is necessary because the OPE of two $U(z)$'s gives
$\Lambda_7(w)$ which can only be cancelled by the terms $\Lambda_7(w)$
coming from $T(z)\Lambda_1(w)$ and $\Lambda_1(z)\Lambda_1(w)$.  As we
will show later, the coefficient $a_0$ of this term is determined to be a
pure number and so can never be tuned to zero  we have some
freedom to adjust other coefficients. (In a different basis where
the $WB_2$ algebra is quadratic, there is no need for $\Lambda_1$
in the BRST operator. Similar remarks also apply to $W_4$.)

For the $W_4$ algebra one can perform similar analysis.
As we noted in ref. [\ZHUOVER], there is a selection rule for $W_4$
algebra. All the fields are classified into even and odd sets.
The even set consists of $T(z)$, $U(z)$, $\Lambda_1(z)$, $\Lambda_2(z)$,
$\Lambda_5(z)$ to $\Lambda_7(z)$ and all their derivatives. The odd set
consists of $W(z)$, $\Lambda_3(z)$, $\Lambda_4(z)$ and all their derivatives.
The OPEs of $(\hbox{even})\times (\hbox{even})$ and
$(\hbox{odd})\times (\hbox{odd})$ give only even fields and the OPEs of
$(\hbox{even})\times (\hbox{odd})$  give only odd fields as one can see from
eqs. $\opewb$ to $\opewl$. If we also assign a parity to the ghost antighost
fields and their derivatives ($(\gamma$, $\beta)$ are odd, $(c$, $b)$ and
$(\delta$, $\alpha)$ are even), the BRST operator or the current is then an
even object.  Bearing this in mind we arrived at the following ansatz for
the BRST current $j(z)$:
$$ j = j_0 + j_1 + j_2 + j_3 , \eqn\jjabcd$$
where the single ghost term $j_0$ is
$$ j_0 = c T + \gamma W + \delta U, \eqn\jja$$
and the three ghost antighost term $j_1$ is
$$ \eqalign{
j_1 = & ~~\big( a_1\gamma \delta' \beta + a_2 \gamma'\delta \beta +
a_3 \gamma\delta\beta' + a_4 \delta\delta^{(3)} b +
a_5 \delta'\delta'' b + a_6 \delta\delta'' b' \cr
& + a_7 \delta\delta' b'' + a_8 \delta\delta'\alpha + a_9 \gamma\gamma' b +
a_{10} c \delta' b + a_{11} c'\delta b + a_{12} c \delta b' \big) T \cr
& + \big( a_{22} \gamma \delta' b + a_{23} \gamma' \delta b + a_{24} \gamma
\delta b' + a_{25} \delta\delta'\beta \big) W +
\big( a_{26} U + a_{27} \Lambda_1 \big) \delta\delta' b \cr
& + cc' b + ( c\gamma' - 2 c' \gamma ) \beta + ( c\delta' - 3 c' \delta)
\alpha + ( 2 c_1 \gamma\gamma^{(3)} - 3 c_2 \gamma'\gamma'') b \cr
& + c_3 \gamma\gamma'\alpha + ( c_4 \gamma\delta^{(3)} - 3 c_5 \gamma'
\delta'' + 5 c_6 \gamma''\delta' - 5 c_7 \gamma^{(3)} \delta ) \beta \cr
& + ( 3 c_8 \delta\delta^{(5)} - 5 c_9 \delta'\delta^{(4)} +
6 c_{10} \delta''\delta^{(3)} ) b +
( c_{11} \delta\delta^{(3)} - 2 c_{12} \delta'\delta'' ) \alpha \cr
& + ( c_{13} c \delta^{(3)} + c_{14}c'\delta'' + c_{15} c'' \delta'
+ c_{16} c^{ (3) } \delta ) b , } \eqn\jjb$$
The five ghost antighost term $j_2$ contains more than 60 terms:
$$ \eqalign{ j_2 =& ~~
\big( a_{13} c \gamma \delta b \beta + a_{14} c \delta\delta' b b'
+ a_{15} \delta\delta'\delta'' b b' + a_{16} \gamma \gamma'
\delta b b' + a_{17} \gamma'\delta\delta' b \beta \cr
& + a_{18} \gamma\delta\delta'' b \beta + a_{19} \gamma \delta\delta' b'
\beta +  a_{20} \gamma\delta\delta' b \beta' \big) T
+ a_{21} \gamma\delta\delta'
b b' W  \cr
& + ( m_8 c' \delta + m_9 \gamma\gamma') c b b'
+ (m_{10} c \delta'\delta'' + m_{11} c \delta\delta^{(3)} +
m_{12} c' \delta\delta'' + m_{13} c'' \delta\delta' ) b b' \cr
& + (m_{14} \gamma\gamma^{(3)} \delta + m_{15} \gamma' \gamma'' \delta
+ m_{16} \gamma\gamma'' \delta' + m_{17} \gamma\gamma'\delta'' ) b b' +
( m_{18} \delta\delta'\delta^{(4)} + m_{19} \delta\delta''\delta^{(3)} )
b b' \cr
& + ( m_{20} c \delta\delta'' + m_{21} c'\delta\delta' ) b \alpha
+ ( m_{22} \gamma\gamma'' \delta + m_{23} \gamma\gamma' \delta')
b \alpha + \delta\delta' ( m_{24} \delta^{(3)} b \alpha  +
m_{25} \delta'' b \alpha' )  \cr
& + c \delta\delta' ( m_{26} b' b'' + m_{27} b \alpha' + m_{28} \beta\beta' ) +
m_{29} \gamma\gamma'\delta b' b'' + \gamma\gamma'
\delta( m_{30} b \alpha' + m_{31} \beta\beta' ) \cr
& + \delta\delta'\delta'' ( m_{32} b' b'' + m_{33} \beta \beta' )
+ (m_{34} c \gamma \delta'' + m_{35} c \gamma'\delta' +
m_{36} c' \gamma \delta' + m_{37} c \gamma'' \delta +
m_{38} c'\gamma' \delta  \cr
& + m_{39} c'' \gamma\delta ) b \beta
+ (m_{40} c \gamma \delta' + m_{41} c \gamma' \delta +
m_{42} c' \gamma \delta ) b \beta' + m_{43} c \gamma\delta b \beta''
+ m_{44} c \gamma \delta \beta\alpha  \cr
& + (
  m_{45} \gamma\delta\delta^{(4)} + m_{46} \gamma\delta'\delta^{(3)}
+ m_{47} \gamma'\delta\delta^{(3)} + m_{48} \gamma'\delta'\delta''
+ m_{49} \gamma''\delta\delta'' + m_{50} \gamma^{(3)}\delta\delta' )
b \beta \cr
& + (
  m_{51} \gamma\delta\delta^{(3)} + m_{52} \gamma\delta'\delta''
+ m_{53} \gamma'\delta\delta'' + m_{54} \gamma''\delta\delta' ) b \beta'
+ m_{55} \gamma\delta\delta' b \beta^{(3)} \cr
& + (
m_{56} \gamma\delta\delta'' + m_{57} \gamma'\delta\delta' ) b \beta''
+ ( m_{58} \gamma\delta\delta'' + m_{59} \gamma'\delta\delta' ) \beta
\alpha  + m_{60} \gamma\delta\delta' \beta\alpha',   }\eqn\jjc $$
and there also exist some 7 ghost antighost terms
$$ \eqalign{ j_3 = & ~~
( m_{61} c \gamma\delta\delta'' + m_{62} c \gamma'\delta\delta'
+ m_{63} c'\gamma\delta\delta' ) b b' \beta + m_{64} c \gamma\delta\delta'
b b' \beta' \cr
& + m_{65} c \gamma\delta\delta' b \beta\alpha +
 \gamma\gamma'\delta\delta'b ( m_{66} b' \alpha + m_{67} \beta\beta' )
+ m_{ 68} \gamma\delta\delta'\delta'' b \beta \alpha \cr
& + ( m_{69} \gamma'\delta\delta'\delta'' + m_{70} \gamma\delta\delta'
\delta^{(3)} ) b b' \beta + m_{71} \gamma\gamma'\delta\delta' b b' b'' +
m_{72} \gamma\delta\delta'\delta'' b b' \beta'  . } \eqn\jjd $$
Fortunately no higher than 7 ghost antighost terms exist. In the next section
we will solve the nilpotent condition $Q^2 = 0$ of such constructed BRST
operators.

\vfill\eject

\centerline{ \bf 4. The Solutions }

\vglue .5cm

The BRST operator is the contour integration of the current $j(z)$:
$$ Q = \oint_0 [\dd z] j(z) . \eqn\papx $$
The square of $Q$ is
$$ Q^2 ={ 1\over 2} \{ Q,Q\} = \oint_0[\dd w] \oint_w[ \dd z] j(z)j(w) .
\eqn\papy$$
To compute $Q^2$ one must do these two integration over $z$ and $w$.
The integration over $z$ is straightforward. Firstly one expands $j(z)j(w)$
and compute the OPEs (or equivalently arranging them into normal ordered
products by doing various contractions). For the $WB_2$ algebra the needed
OPEs for the bosonic fields
$T(z)$, $U(z)$ and $\Lambda_1(z)$ are given in $\opewb$, $\opea$ and
$\opeb$. Notice that we have purposely expanded these OPEs up to the fields
with highest spin 6. This is the necessary and sufficient degree of accuracy.
Because of the anti-commutativity of ghost antighost fields, terms containing
higher spin
bosonic fields are automatically zero. (One can easily convince oneself by
constructing the lowest spin product of ghost antighost fields with ghost
number 2.) For  the contraction
of ghost antighost fields we use exactly Wick theorem theorem in quantum
field theory while taking into account the fermionic property of these
(free) fields. After doing all these, the integration over $z$ amounts to
evaluate the residue of the resulting expression at $w$.

The integration over $w$ is actually not needed because our purpose
is to set $Q^2$ to zero and determine all the unknown coefficients.
Evidently the integrand should be a total derivative if the contour
integration is zero. This is necessary and sufficient.
By setting the integrand to total derivatives one obtains a set of
equations among the unknown coefficients and the central charge $c$.
All these equation can have a  solution only if the central charge is
$c = 172$. This is in agreement with the
simple counting that the total central charge of the matter and ghost
antighost system is zero. (The $(b$, $c)$ system has central charge $-26$ and
the $(\alpha, \delta)$ system has central charge
$- 146$.) By setting $c= 172$ and solving these equations
we found that some of the coefficients are pure numbers
$$ a_0 = { 253 \over 654444}, \qquad m_4 = 4 , \qquad
m_5 = 3 , \qquad  m_6 = 1 .  \eqn\coeb $$
The rest coefficients depend on three arbitrary parameters which
we denoted as $c_i$ $(i=1$, 2, 3). We have (in order of increasing
complexity)
$$ \eqalign{ &
{
\eqalign{
a_6 = & ~~ c_{1}\cr
a_9 = & ~~ c_{2}\cr
m_{20} = & -7\,c_{1}\cr
a_1 = & ~~ {1\over {21\,{\sqrt{742}}}} - c_{2}\cr
m_9 = & ~~ {{11\,c_{1}}\over 2} + {{17\,c_{2}}\over 2}\cr }
\qquad
\eqalign{
a_7 = & ~~ 2 \,c_{1}\cr
m_{13} = & ~~ c_{1}\cr
m_{21} = & -4\,c_{1}\cr
m_7 = & ~~ {{13\,c_{1}}\over 2} + 12\,c_{2}\cr
m_{10} = & ~~ {{5\,c_{1}}\over 2} + {{5\,c_{2}}\over 3}\cr }
\qquad
\eqalign{
a_8 = & ~~ c_{1}\cr
m_{19} = & ~~ 4\,c_{1}\cr
m_{11} = & {{-7\,c_{2}} / 2}\cr
m_{12} = & ~~ {{7\,c_{2}}\over 2}\cr
m_8 = & - {{2\,c_{1}}\over 3} + {{31\,c_{2}}\over 6}\cr }
 }\cr
& \cr
&{
\eqalign{
a_2 = & ~~ {{337}\over {311640}} + {{7\,{{c_{1}}^2}}\over {60}} +
   {{71\,c_{1}\,c_{2}}\over {60}} - {{41\,{{c_{2}}^2}}\over {60}} -
   {{c_{3}}\over 5}\cr
a_3 = & -{{239}\over {155820}} + {{{{c_{1}}^2}}\over 5} +
   {{71\,c_{1}\,c_{2}}\over {60}} + {{2\,{{c_{2}}^2}}\over 5}
- {{c_{3}}\over 5}\cr
a_4 = & ~~ {1\over {1060}} - {{17\,{{c_{1}}^2}}\over {20}} +
   {{71\,c_{1}\,c_{2}}\over {30}} - {{29\,{{c_{2}}^2}}\over {20}} -
   {{2\,c_{3}}\over 5}\cr
a_5 = & ~~ {1\over {1060}} - {{7\,{{c_{1}}^2}}\over 4} -
   {{3\,{{c_{2}}^2}}\over 4} \qquad
a_{11} =  {{c_{1}}^2} + c_{1}\,c_{2} \cr
m_1 = & ~~ {{43}\over {63600}} -
   {{127\,{{c_{1}}^2}}\over {100}} + {{3037\,c_{1}\,c_{2}}\over {1800}} -
   {{29\,{{c_{2}}^2}}\over {600}} + {{14\,c_{3}}\over {75}}\cr
m_2 = & -{{43}\over {38160}} + {{149\,{{c_{1}}^2}}\over {120}} +
   {{217\,c_{1}\,c_{2}}\over {720}} + {{81\,{{c_{2}}^2}}\over {40}} +
   {{13\,c_{3}}\over {60}}\cr
m_3 = & ~~ {{43}\over {44520}} +
   {{77\,{{c_{1}}^2}}\over {30}} + {{497\,c_{1}\,c_{2}}\over {360}} -
   {{17\,{{c_{2}}^2}}\over {60}} - {{7\,c_{3}}\over {30}}\cr
m_{14} = & ~~ {{43}\over {51940}} - {{14\,{{c_{1}}^2}}\over 5} +
   {{781\,c_{1}\,c_{2}}\over {60}} - {{151\,{{c_{2}}^2}}\over {10}} -
   {{11\,c_{3}}\over 5}\cr
m_{15} = & ~~ {{43}\over {77910}} - {{58\,{{c_{1}}^2}}\over {15}} +
   {{71\,c_{1}\,c_{2}}\over {15}} - {{151\,{{c_{2}}^2}}\over {15}} -
   {{4\,c_{3}}\over 5}\cr
m_{22} = & ~~ {{5\,{{c_{1}}^2}}\over 2} +
   {{155\,c_{1}\,c_{2}}\over {12}} + c_{3} \qquad
m_{23} = - {{5\,{{c_{1}}^2}}\over 2} -
   {{155\,c_{1}\,c_{2}}\over {12}} + c_{3}\cr
m_{24} = & ~~ 3\,{{c_{1}}^2} + {{5\,c_{1}\,c_{2}}\over 3} + 2\,c_{3}\quad
m_{25} =  {{17\,{{c_{1}}^2}}\over 6}
   + {{3\,c_{1}\,c_{2}}\over 4} + c_{3}\quad
m_{26} =  - {{215\,c_{1}\,c_{2}}\over {12}} + c_{3}\cr
a_{10} = & - {{1}\over {2646\,{\sqrt{742}}}} + {{239\,c_{1}}\over {155820}} +
   {{{{c_{1}}^3}}\over {30}} + {{163\,c_{2}}\over {20776}}
  +{{13\,{{c_{1}}^2}\,c_{2}}\over {30}}
  -{{11\,c_{1}\,{{c_{2}}^2}}\over {10}} -
   {{13\,{{c_{2}}^3}}\over {12}} - {{c_{1}\,c_{3}}\over 5}\cr   }
} 
 }  $$

$$ \eqalign{
m_{16} = & ~~ {1\over {3969\,{\sqrt{742}}}} +
   {{1849\,c_{1}}\over {1869840}} - {{899\,{{c_{1}}^3}}\over {360}} -
   {{473\,c_{2}}\over {93492}} \cr
& - {{1907\,{{c_{1}}^2}\,c_{2}}\over {360}} -
   {{217\,c_{1}\,{{c_{2}}^2}}\over {40}} + {{199\,{{c_{2}}^3}}\over {18}} +
   {{67\,c_{1}\,c_{3}}\over {30}} + c_{2}\,c_{3}\cr
m_{17} = & ~~ {5\over {5292\,{\sqrt{742}}}} - {{43\,c_{1}}\over {38955}} -
   {{103\,{{c_{1}}^3}}\over {30}} - {{258\,c_{2}}\over {12985}} \cr
&  - {{799\,{{c_{1}}^2}\,c_{2}}\over {90}} -
   {{2659\,c_{1}\,{{c_{2}}^2}}\over {90}} + {{391\,{{c_{2}}^3}}\over {15}} +
   {{49\,c_{1}\,c_{3}}\over {15}} + {{82\,c_{2}\,c_{3}}\over {15}}\cr
m_{18} = & - {{11}\over {15876\,{\sqrt{742}}}} +
   {{145\,{{c_{1}}^3}}\over {36}} +
   {{13373\,c_{2}}\over {934920}} + {{1063\,{{c_{1}}^2}\,c_{2}}\over {120}}\cr
&+{{994\,c_{1}\,{{c_{2}}^2}}\over {45}} - {{3581\,{{c_{2}}^3}}\over {180}} -
   {{7\,c_{1}\,c_{3}}\over 2} - {{56\,c_{2}\,c_{3}}\over {15}}\cr }$$
With the above explicit expressions for all these coefficients, the BRST
operator is shown to be nilpotent for arbitrary $c_{i}$'s. We will discuss
this solution in the next section.

Quite similarly but becoming more complicated,
the above procedure can be carried out for the $W_4$ algebra.
The central charge must be $246$ in order to have a nilpotent BRST operator.
This is the number we would expect for the ghost antighost system $(b$, $c)$,
$(\beta$, $\gamma)$ and $(\alpha$, $\delta)$. However the explicit expression
of all these coefficients are quite long. Let us first state the results in
words. All the coefficients are determined as algebraic functions of seven
coefficients which we choose (quite arbitrarily) to be $c_{15}$, $c_{16}$,
$m_8$, $m_{14}$, $m_{22}$, $m_{29}$ and $m_{32}$. The explcit expression of
all the coefficients in terms of these seven (free) coefficients are quite
long. Nevertheless, with these explicit expressions we do succeed in proving
that the BRST operator is indeed nilpotent.

Having stated the result in general terms let us be more specific.
As we will discuss in more details for the BRST operator of the $WB_2$
algebra in the next section, a quite natural procedure to fix these free
coefficients is to require $\{Q, b(z)\}$ to be the total stress-energy tensor.
By imposing this condition, five out of the seven free coefficients are fixed.
Quite surprisingly the resulting expression for all the coefficients become
quite simple. 36 coefficients are identically equal to zero. These are:
$a_{10}$ to $a_{14}$, $c_{13}$ to $c_{16}$, $m_{8}$ to $m_{13}$,
$m_{20}$ and $m_{21}$, $m_{25}$ to $m_{27}$, $m_{34}$ to $m_{44}$ and $m_{61}$
to $m_{64}$. Some coefficients are pure numbers:
$$\eqalign{
c_3 = & ~~2  \cr
a_{27} = &~~ {{139909}\over {2449225}} \cr } \qquad
\eqalign{
a_9 = & ~~{{506}\over {1565}} \cr
c_{11} = & -{{1703}\over {6260}} \cr } \qquad
\eqalign{
a_{25} = &~~ {3\over 8} \cr
m_{33} = & -{{69}\over {64}} \cr } $$
Setting $m_{22} = cv$ and trading the other free parameters as $vc$ (which is
identified with the free parameter $\beta$ in ref. [\HORN]) the rest
coefficients are
$$
\eqalign{
a_1 =  & ~~{{37\,{\it cv}}\over {26}} \cr
a_7 = & ~~{{258819}\over{1252000}}-{{143301\,{{{\it cv}}^2}}\over {66248}} \cr
a_{22} = &~~ {{553}\over {1565}} - {{106\,{\it cv}}\over {91}} \cr
a_{26} = & -{{459}\over {1565}} + {{185\,{\it cv}}\over {182}} \cr
c_4 = & ~~{{2737}\over {12520}} + {{6\,{\it cv}}\over {91}} \cr
c_7 = & ~~{{9891}\over {62600}} + {{20\,{\it cv}}\over {91}} \cr
m_{30} = & ~~{{1518}\over {1565}} - {{459\,{\it cv}}\over {91}} \cr
m_{58} = & -{{918}\over {1565}} + {{279\,{\it cv}}\over {91}} \cr  }\quad
\eqalign{
a_2 =  & - {{159\,{\it cv}}\over {91}} \cr
a_8 = & -{{185\,{\it cv}}\over {182}} \cr
a_{23} = & -{{1059}\over {1565}} + {{365\,{\it cv}}\over {182}} \cr
c_1 = & ~~{{12397}\over {18780}} - {{1271\,{\it cv}}\over {1092}} \cr
c_5 = & -{{1547}\over {12520}} + {{675\,{\it cv}}\over {364}} \cr
c_{12} = & ~~{{1013}\over {25040}} - {{185\,{\it cv}}\over {104}} \cr
m_{31} = & -{{1941}\over {6260}} + {{495\,{\it cv}}\over {182}} \cr
m_{59} = & -{{3777}\over {3130}} + {{1275\,{\it cv}}\over {182}} \cr }\quad
\eqalign{
a_3 = & ~~{{47\,{\it cv}}\over {182}} \cr
a_{16} = & {{512072}\over{2449225}}-{{116127\,{\it cv}}\over {142415}} \cr
a_{24} = & -{{459}\over {3130}} + {{47\,{\it cv}}\over {182}} \cr
c_2 = & ~~{{12397}\over {18780}} - {{159\,{\it cv}}\over {91}} \cr
c_6 = & ~~{{832}\over {7825}} + {{35\,{\it cv}}\over {52}} \cr
m_{23} = &~~ {{2024}\over {1565}} - {{703\,{\it cv}}\over {91}} \cr
m_{25} = & -{{15931409}\over {1773553152}} - 11\,{\it vc} \cr
m_{60} = & -{{8367}\over {6260}} + {{1335\,{\it cv}}\over {182}} \cr } $$
and
$$\eqalign{
 a_4 = & ~~{{145947672356389}\over {763292937792000}} + {\it vc} +
  {{621693\,{\it cv}}\over {2506504}} -
  {{1060309\,{{{\it cv}}^2}}\over {1093092}} \cr
 a_5 = & -{{355891353306971}\over {763292937792000}} + {\it vc} +
   {{621693\,{\it cv}}\over {2506504}} +
   {{1698633\,{{{\it cv}}^2}}\over {728728}} \cr
 a_6 = & ~~ {{48206198862109}\over {381646468896000}} + 2\,{\it vc} +
   {{621693\,{\it cv}}\over {1253252}} -
   {{222333\,{{{\it cv}}^2}}\over {104104}} \cr
 a_{17} = & ~~{{80986927589317}\over {381646468896000}} + 2\,{\it vc} +
   {{621693\,{\it cv}}\over {1253252}} -
   {{2697385\,{{{\it cv}}^2}}\over {364364}} \cr
 a_{18} = & -{{76962564246203}\over {381646468896000}} + 2\,{\it vc} +
   {{621693\,{\it cv}}\over {1253252}} +
   {{793465\,{{{\it cv}}^2}}\over {364364}} \cr
 a_{19} = & -{{30689454039803}\over {190823234448000}} + 4\,{\it vc} +
   {{621693\,{\it cv}}\over {626626}} +
   {{4995\,{{{\it cv}}^2}}\over {91091}} \cr
 a_{20} = & -{{33360383297723}\over {381646468896000}} + 2\,{\it vc} +
   {{621693\,{\it cv}}\over {1253252}} -
   {{489135\,{{{\it cv}}^2}}\over {364364}} \cr
 a_{21} = & ~~ {{4209291047557}\over {381646468896000}} + 2\,{\it vc} -
   {{32547\,{\it cv}}\over {89518}} +
   {{123789\,{{{\it cv}}^2}}\over {91091}} \cr
 c_8 = & ~~{{38110049205397}\over {490688317152000}} -
   {{14\,{\it vc}}\over {45}} -
   {{1537379\,{\it cv}}\over {10742160}} -
   {{534983\,{{{\it cv}}^2}}\over {1093092}} \cr
 c_9 = & ~~{{52763531459717}\over {366380610140160}} +
   {{13\,{\it vc}}\over {60}} -
   {{371291\,{\it cv}}\over {2148432}} -
   {{6567019\,{{{\it cv}}^2}}\over {8744736}} \cr
 c_{10} = & ~~{{26716946702321}\over {785101307443200}} +
   {{7\,{\it vc}}\over {36}} +
   {{52087\,{\it cv}}\over {661056}} -
   {{712139\,{{{\it cv}}^2}}\over {8744736}} \cr }$$

$$\eqalign{
 m_{14} = & ~~{{206930376169691}\over {163562772384000}} -
   {{14\,{\it vc}}\over 3} - {{113136277\,{\it cv}}\over {18798780}} +
   {{938965\,{{{\it cv}}^2}}\over {182182}} \cr
 m_{15} = & -{{87631707259531}\over {25443097926400}} - 2\,{\it vc} +
   {{156384181\,{\it cv}}\over {7519512}} -
   {{5820277\,{{{\it cv}}^2}}\over {182182}} \cr
 m_{16} = & ~~{{666084599304911}\over {381646468896000}} + 6\,{\it vc} -
   {{3226031\,{\it cv}}\over {289212}} +
   {{3572671\,{{{\it cv}}^2}}\over {182182}} \cr
 m_{17} = & -{{86745615934751}\over {76329293779200}} + 2\,{\it vc} +
   {{80929421\,{\it cv}}\over {9399390}}
  - {{3112537\,{{{\it cv}}^2}}\over {182182}} \cr
 m_{24} = & ~~{{2368726689487}\over {38164646889600}} - 4\,{\it vc} -
   {{4433323\,{\it cv}}\over {5013008}} +
   {{271025\,{{{\it cv}}^2}}\over {56056}} \cr
 m_{29} = & -{{125960351}\over {117562800}} +
   {{16960361\,{\it cv}}\over {3417960}} -
   {{12301\,{{{\it cv}}^2}}\over {2366}} \cr
 m_{45} = & -{{1823102103976543}\over {4579757626752000}} -
   {{19\,{\it vc}}\over 6} + {{79769\,{\it cv}}\over {447590}} +
   {{175935\,{{{\it cv}}^2}}\over {112112}} \cr
 m_{46} = & -{{201321547925903}\over {143117425836000}} +
   {{16\,{\it vc}}\over 3} +
   {{7622399\,{\it cv}}\over {1790360}} +
   {{922225\,{{{\it cv}}^2}}\over {728728}} \cr
 m_{47} = & ~~{{523287237}\over {391876000}} -
   {{10763643\,{\it cv}}\over {2278640}} -
   {{264207\,{{{\it cv}}^2}}\over {66248}} \cr
 m_{48} = & ~~{{11520872866549}\over {109041848256000}} + 7\,{\it vc} -
   {{7651269\,{\it cv}}\over {12532520}} +
   {{4241739\,{{{\it cv}}^2}}\over {364364}} \cr
 m_{49} = & -{{40529012472799}\over {152658587558400}} + {\it vc} -
   {{8187009\,{\it cv}}\over {25065040}} +
   {{4962\,{{{\it cv}}^2}}\over {1001}} \cr
 m_{50} = & ~~{{136142989370639}\over {114493940668800}} -
   {{20\,{\it vc}}\over 3} -
   {{13780917\,{\it cv}}\over {2506504}} +
   {{136075\,{{{\it cv}}^2}}\over {728728}} \cr
 m_{51} = & -{{1001912494309861}\over {763292937792000}} - {\it vc} +
   {{9178583\,{\it cv}}\over {6266260}} +
   {{635931\,{{{\it cv}}^2}}\over {364364}} \cr
 m_{52} = & -{{373185538120107}\over {127215489632000}} + 6\,{\it vc} +
   {{132646001\,{\it cv}}\over {12532520}} -
   {{295651\,{{{\it cv}}^2}}\over {91091}} \cr
 m_{53} = & ~~{{239563804445407}\over {95411617224000}} + 8\,{\it vc} -
   {{260289693\,{\it cv}}\over {25065040}} +
   {{1570815\,{{{\it cv}}^2}}\over {182182}} \cr
 m_{54} = & ~~{{113975631706187}\over {152658587558400}} - 5\,{\it vc} -
   {{30290837\,{\it cv}}\over {5013008}} +
   {{6156635\,{{{\it cv}}^2}}\over {728728}} \cr
 m_{55} = & -{{130048022434427}\over {457975762675200}} +
   {{5\,{\it vc}}\over 3} -
   {{1336781\,{\it cv}}\over {1253252}} +
   {{16525\,{{{\it cv}}^2}}\over {4004}}  \cr
m_{56} = & -{{63541384149247}\over {38164646889600}} + 4\,{\it vc} +
   {{17699379\,{\it cv}}\over {6266260}} +
   {{467931\,{{{\it cv}}^2}}\over {182182}} \cr
 m_{57} = & ~~{{12043053}\over {4898450}} -
   {{3067341\,{\it cv}}\over {227864}} +
   {{607545\,{{{\it cv}}^2}}\over {33124}} \cr
 m_{66} = & -{{70585843840763}\over {95411617224000}} + 8\,{\it vc} +
   {{8732908\,{\it cv}}\over {1566565}} -
   {{962718\,{{{\it cv}}^2}}\over {91091}} \cr  }$$

$$\eqalign{
 m_{67} = & ~~{{73757449136699}\over {76329293779200}} - 10\,{\it vc} -
   {{12120831\,{\it cv}}\over {1253252}} +
   {{8735145\,{{{\it cv}}^2}}\over {364364}} \cr
 m_{68} = & -{{15583656166597}\over {95411617224000}} - 8\,{\it vc} +
   {{13299969\,{\it cv}}\over {6266260}} -
   {{162615\,{{{\it cv}}^2}}\over {14014}} \cr
a_{15} = & ~~{{2096636806}\over {3833037125}} -
   {{143897097259880989\,{\it cv}}\over {46306438226048000}} +
   {{789\,{\it vc}\,{\it cv}}\over {182}} \cr
   &  + {{490515777\,{{{\it cv}}^2}}\over {456183728}} +
   {{395882425\,{{{\it cv}}^3}}\over {33157124}} \cr
m_{18} = & ~~{{11126358008106501049}\over {11467713097387008000}} +
   {{31333\,{\it vc}}\over {15024}} -
   {{2300259221256327283\,{\it cv}}\over {333406355227545600}} \cr
&  + {{145\,{\it vc}\,{\it cv}}\over {168}} +
   {{206953330853\,{{{\it cv}}^2}}\over {18247349120}} +
   {{507905105\,{{{\it cv}}^3}}\over {122426304}} \cr
m_{32} = & -{{4320751301911}\over {2943772512000}} +
   {{14751898953\,{\it cv}}\over {1426428640}} -
   {{133780861\,{{{\it cv}}^2}}\over {7405580}} -
   {{6840745\,{{{\it cv}}^3}}\over {10334688}} \cr
m_{19} = & -{{212279573026507577}\over {273040788033024000}} +
   {{2205\,{\it vc}}\over {2504}} +
   {{224455082507749573\,{\it cv}}\over {46306438226048000}} \cr
& - {{933\,{\it vc}\,{\it cv}}\over {182}} -
   {{28655403533\,{{{\it cv}}^2}}\over {4561837280}} -
   {{3964530205\,{{{\it cv}}^3}}\over {795770976}} \cr
m_{69} = & ~~ {{2934212453988242167}\over {1592737930192640000}} +
   {{12513\,{\it vc}}\over {6260}} -
   {{191792451947131327\,{\it cv}}\over {11576609556512000}} \cr
& -{{618\,{\it vc}\,{\it cv}}\over {91}} +
   {{253409589117\,{{{\it cv}}^2}}\over {4561837280}} -
   {{1364553837\,{{{\it cv}}^3}}\over {18946928}} \cr
m_{70} = & -{{957021797194575577}\over {2389106895288960000}} +
   {{21179\,{\it vc}}\over {3130}} +
   {{39163300881728239\,{\it cv}}\over {27783862935628800}} \cr
 & - {{1489\,{\it vc}\,{\it cv}}\over {182}} +
   {{186796629\,{{{\it cv}}^2}}\over {912367456}} -
   {{17797111\,{{{\it cv}}^3}}\over {33157124}} \cr
m_{71} = & -{{37275038049030721}\over {72397178645120000}} +
   {{52371\,{\it vc}}\over {3130}} +
   {{790319105345479\,{\it cv}}\over {485731869504000}} \cr
 & - {{3883\,{\it vc}\,{\it cv}}\over {91}} +
   {{20276331\,{{{\it cv}}^2}}\over {3190096}} -
   {{59350407\,{{{\it cv}}^3}}\over {3014284}} \cr
m_{72} = & ~~{{8572847559656227}\over {20076528531840000}} +
   {{46529\,{\it vc}}\over {3130}} -
   {{51941209398171283\,{\it cv}}\over {6314514303552000}} \cr
 & - {{1901\,{\it vc}\,{\it cv}}\over {91}} +
   {{10675480773\,{{{\it cv}}^2}}\over {285114830}} -
   {{3255147397\,{{{\it cv}}^3}}\over {66314248}}  } $$
In the next section after we have a better understanding of the connection
between these free parameters and the canonical transformation of the ghost
antighost fields, we will show how one can obtain the complete solution
from the above special solution.

\vfill\eject

\centerline{\bf  5. Discussions }
\vglue .5cm

In this section we will analyse the solutions obtained in the last section
in details and explicitly show that the free parameters in the BRST
operator are related to the canonical transformation of the ghost antighost
fields as first discussed in ref. [\HORN] for the BRST operator of the $W_4$
algebra.

First let us look at the BRST operator of the $WB_2$ algebra. As in other
extended conformal algebras, we define a generalized (total) stress-energy
tensor
$$ \eqalign{ T_{\hbox{ tot. }} (z)  \equiv & ~~ \{ Q , b(z) \} \cr
= & ~~ T(z) + 2 c'(z) b(z) + c(z)b'(z) + 4 \delta'(z) \alpha(z) +
3 \delta(z) \alpha'(z) + (a_6 b\delta' + a_7 b'\delta) T \cr
& - a_8 (b\delta T)' + a_{11}bb'\delta
\delta' T + m_7 b'\delta^{(3)} + m_8 b \delta^{(3)} + m_9 b''\delta'
+ m_{10} b^{(3)} \delta \cr
& + a_{11} b b'\delta\delta' T + m_{13}\big(
c'bb'\delta + (cbb'\delta)' \big) + \big( m_{19} b'\delta'\alpha
+m_{20} b\delta'\alpha' + m_{21} b \delta''\alpha\big) \delta \cr
& + \big( m_{22} bb' \delta \delta^{(3)} +
m_{23} bb' \delta'\delta''  + m_{24} bb'' \delta\delta''
+ m_{25} bb^{(3)}  \delta \delta' + m_{26} b'b'' \delta\delta' \big) . }
\eqn\wwa $$
As one can see from the above, this stress-energy tensor contains
more terms than needed (we need only the terms on the second line of the
above equation). So a natural choice for the free parameters $c_{i}$'s
is to require these extra terms to be zero. This fixes all the $c_i$'s to
zero. With these zero $c_i$, all the coefficients become pure numbers
and we have a unique BRST operator. This BRST operator is the contour
integration of the following BRST current
$$\eqalign{
j = &  ~~ c(z)T(z) + \delta(z) U(z) +  b\delta\delta'
\Big( {253 \over 65444}  \Lambda_1 + {1\over 21\sqrt{742}}  U\Big) \cr
& +{ 1\over 1060} \Big( b' \delta\delta''  +  b'' \delta \delta'
+ {337 \over 294} b\delta\delta^{(3)}
- {239\over 147}  b \delta' \delta''  \Big) T \cr
& + {43 \over 636} \Big(
{1\over 10} b \delta\delta^{(5)}  -{1\over 6}b \delta' \delta^{(4)}
+{1\over 7} b \delta'' \delta^{(3)}  \Big)  + c \big( 4 \delta' \alpha
+3 \delta\alpha' \big) + c c' b  \cr
& -{1\over 2646\sqrt{742} } cb b' \delta\delta' T
+ {43 \over 25970} \Big(
{1\over 2} b\delta \delta' \delta'' \alpha'
+ {1\over 3}b\delta \delta'\delta^{(3)} \alpha \Big) \cr
& +{1\over 1323\sqrt{742} } \Big(
{1\over 3}  b b'\delta \delta' \delta^{(4)}
+ {5\over 4} b b'\delta \delta''\delta^{(3)}
-{11\over 12} b  b^{(3)}\delta  \delta'\delta'' \Big) . } \eqn\oooo$$

Having obtained such simple expression for the BRST operator of the nonlinear
$WB_2$ algebra, we would like to ask why there appear so many free parameters
in $Q$ and they could all be fixed by requiring that the total stress-energy
tensor is the BRST commutator of the antighost $b(z)$.
As it was first discussed in
ref. [\HORN] for the $W_4$ algebra, these free parameters are related
to the canonical transformation of the ghost antighost fields. Now we show
that this is also true for the $WB_2$ algebra.

If there are only $(b,c)$ and $(\beta,\gamma)$ ghost antighost fields, there
doesn't exist any  canonical transformations (except the parity violating
linear transformation). However for the ghost antighost fields  $(b,c)$ and
$(\alpha,\delta)$ appearing in the BRST quantization  of the $WB_2$ algebra,
there exists a three parameter family of canonical transformation. These
transformations form an Abelian group. By explicit calculation we found
that these transformations are generated from the following three
independent transformations:
$$\eqalign{&\qquad  \big\{ b, c, \alpha, \delta\big\} \longrightarrow \cr
& \left\{ b + c_1\,bb'\delta,
  c - c_1\,\left( cb\delta' + 2\,cb'\delta + c'b\delta \right)  +
   {{{{c_1}^3}}\over 6}\,bb'\delta\delta' \delta'' \right. \cr
& ~~  + {{{{c_1}^2} }\over {12}}\,\left( 12\,cbb'\delta\delta'
-b\delta\delta^{(3)} - 15\,b'\delta\delta'' -
         21\,b''\delta\delta'  \right),\alpha - c_1\,cbb' \cr
& ~~ \left.  + {{{{c_1}^2} }\over {12}}\,\left( 6\,b'b''\delta
-9\,bb'\delta'' - 15\,bb''\delta' -
         7\,bb^{(3)}\delta  \right) +
   {{13\,{{c_1}^3}}\over {12}}\,bb'b''\delta \delta',\delta \right\} , \cr
& \cr
&\left\{  b,c - c_2\,\delta\delta'\alpha -
   {{{{c_2}^2} }\over {12}}\,\left( 13\,b\delta\delta^{(3)} +
         27\,b'\delta\delta'' + 9\,b''\delta\delta' \right) +
   {{13\,{{c_2}^3}}\over {12}}\,bb'\delta\delta' \delta'', \right. \cr
& ~~ \alpha + c_2\,\left( b\delta\alpha' + 2\,b\delta'\alpha +
      b'\delta\alpha \right)
- {{{{c_2}^3}}\over 3}\, bb'b''\delta\delta' \cr
& ~~ \left.  + {{{{c_2}^2}}\over {12}}
\,\left( 5\,bb^{(3)}\delta -21\,bb'\delta'' -
         3\,bb''\delta'  +
         6\,b'b''\delta + 12\,bb'\delta\delta'\alpha \right) ,
\delta + c_2\,b\delta\delta' \right\} ,\cr
& \cr
& \left\{  b,c - c_3\,\left( b\delta \delta^{(3)} + b\delta'\delta'' +
      2\,b'\delta\delta'' \right) ,
  \alpha + c_3\,\left( 2\,bb''\delta' + bb^{(3)}\delta +
      b'b''\delta \right) ,\delta\right\} , } \eqn\canaa $$
where $c_1$, $c_2$ and $c_3$ are free parameters of these transformation.
As one can show by explicit calculation (see below), non of the above
transformations or any combination of them leaves the standard ghost
antighost stress-energy tensor $T_{\hbox{gh.}}$, which is given by
$$ T_{ \hbox{gh.}}(z) = 2 c'(z) b(z) + c(z)b'(z) + 4 \delta'(z) \alpha(z) +
3 \delta(z) \alpha'(z) , \eqn\tghaa $$
invariant. This is why we get a unique BRST operator
after imposing the condition: $\{Q, b(z)\} = $ $T(z) + T_{\hbox{gh.}}(z)$.
The general solution can be obtained from this unique one just by doing a
general canonical transformation of these ghost antighost fields. Of course
the reverse is also true. In this way we established the connection between the
free parameters in the BRST operator and the canonical transformation of
the ghost antighost fields for $WB_2$.

Quite similarly but becoming more complicated, the above connection also
applies to the $W_4$ algebra, as first discussed in ref. [\HORN]. Apart from
the three parameter transformations given in eq. $\canaa$, we have four more
independent canonical transformations by adding $(\beta, \gamma)$ to the
existing ghost antighost fields in $WB_2$. Explicitly these four
transformations are ($b$ and $\delta$ are unchanged)
$$\eqalign{ & \qquad \big\{ c, \beta, \gamma, \alpha \big\}
\longrightarrow \cr
& \left\{c + c_4\,\gamma\beta'\delta + {{{{c_4}^2}}\over {12}}
\left(  12b\gamma\beta\delta\delta'' -b\delta\delta^{(3)}
- 3b'\delta\delta'' - 3b''\delta\delta' +
         12b\gamma'\beta\delta\delta' +
         24b'\gamma\beta\delta\delta' \right) \right. \cr
& ~ - {{2{c_4^3}}\over 3}bb'\delta\delta' \delta'',
\beta - c_4b\beta'\delta + {c_4^2}bb'\beta\delta\delta',
  \gamma - c_4\left( b\gamma\delta' + b\gamma'\delta +
      b'\gamma\delta \right)  - {c_4^2}bb'\gamma\delta\delta',
  \alpha - c_4b\gamma\beta' \cr
&~\left. + {{2{{c_4}^3}}\over 3}bb'b''\delta
        \delta' + {{{{c_4}^2} }\over {12}}
 \left(24bb'\gamma\beta\delta' -3bb'\delta'' -
         3bb''\delta' - bb^{(3)}\delta
          + 12bb'\gamma'\beta\delta +
         12bb''\gamma\beta\delta \right)\right\} ,\cr
& \cr
& \left\{  c - c_5\,\gamma'\beta\delta -
   {{{{c_5}^3}}\over 6}\,bb'\delta\delta' \delta'' - {{{{c_5}^2}
       }\over {12}}\,
       \left( b\delta\delta^{(3)} + 3\,b'\delta\delta'' +
         3\,b''\delta\delta' - 12\,b\gamma'\beta\delta\delta' \right)
\right. ,\cr
& ~~ \beta - c_5\, \left( b\beta\delta' +
      b\beta'\delta + b'\beta\delta \right) ,
  \gamma - c_5\,b\gamma'\delta ,\cr & \left. \alpha + c_5\,b\gamma'\beta -
   {{{{c_5}^2}  }\over {12}}\,\left( 3\,bb'\delta'' + 3\,bb''\delta' +
         bb^{(3)}\delta - 12\,bb'\gamma'\beta\delta \right)
      + {{{{c_5}^3}}\over 6}\, bb'b''\delta\delta'
 \right\},  \cr
& \cr
&\left\{  b,c + c_6\,\gamma\beta\delta' -
   {{{{c_6}^2} }\over 2}\,\left( b\delta\delta^{(3)} +
         2\,b'\delta\delta'' \right) ,\beta - c_6\,b\beta\delta',
  \gamma + c_6\,b\gamma\delta' ,\right. \cr
&~~\left. \alpha + c_6\,\left( b\gamma\beta' +
      b\gamma'\beta + b'\gamma\beta \right)  +
   {{{{c_6}^2} }\over 2}\,\left(  bb''\delta' + bb^{(3)}\delta
 + b'b''\delta - bb'\delta''  \right)\right\} , \cr
& \cr
& \left\{ b,c - c_7\,\left( b\gamma\beta
       \delta\delta'' + b\gamma\beta'\delta
       \delta' + b\gamma'\beta\delta \delta' + 2\,b'\gamma\beta\delta
        \delta' \right) ,  \beta - c_7\,bb'\beta\delta\delta', \right. \cr
&~~ \left. \gamma + c_7\,bb'\gamma\delta\delta', \alpha
 - c_7\,\left( 2\,bb'\gamma\beta
        \delta' + bb'\gamma\beta' \delta + bb'\gamma'\beta\delta +
      bb''\gamma\beta\delta \right)  \right\}.  }\eqn\canbb  $$
where $c_4$ to $c_7$ are free parameters of these transformations.
In total we have a seven parameter canonical transformation of
the ghost antighost field in $W_4$. So we should expect at least seven free
parameters in the BRST operator if there exists one BRST operator. This is
indeed the case as we have found precisly a seven parameter family of
nilpotent BRST operator by explicit construction. As one can show by
calculation there exist three independent canonical transformations,
parametrized by $c_4$, $c_5$ and $c_7$ with the other $c_i$'s depending
on them as follows
$$\eqalign{ \left\{ c_1, c_2,c_3,c_6\right\} = &  \Big\{
 {{53\,a_4}\over {64}} - {{47\,a_5}\over {384}},
  {{-23\,a_4}\over {64}} - {{203\,a_5}\over {384}}, \cr
 {{435\,{{a_4}^2}}\over {4096}} & - {{21145\,a_4\,a_5}\over {12288}} -
   {{101045\,{{a_5}^2}}\over {147456}} - {{a_7}\over 2},
  a_4 + {{2\,a_5}\over 3}\Big\}, } $$
which leave the ghost antighost stress-energy tensor invariant.
Nevertheless only two of them also leave $b(z)$ unchanged. This is why
we found only a two (instead of three) parameter family of BRST charge after
imposing the condition:
$$ T_{\hbox{tot.}} \equiv \{Q, b(z)\} = T(z) + T_{\hbox{gh.}}(z),
\eqn\ooooo$$
where
$$ T_{\hbox{gh.}}(z) =  2 c'(z) b(z) + c(z)b'(z)
+ 3 \gamma'(z) \beta(z) + 2 \gamma(z) \beta'(z)
 + 4 \delta'(z) \alpha(z) +
3 \delta(z) \alpha'(z) , \eqn\oook $$
is the stress-energy tensor of the ghost antighost fields in $W_4$.
With eqs. $\canaa$ and $\canbb$ and the explicit solution given in
the last section, one can easily obtain the most general solution
with seven free parameters.

\REF\POPEB{H. Lu, C. N. Pope, S. Schrans and X. J. Wang, On the spectrum
and scattering of $W_3$ strings, preprint CTP-TAMU-4/93 or
KUL-TF-93/2 (January 1993)}
\REF\LAS{E. Bergshoeff, H. J. Boonstra, M. de Roo, S. Panda and A.
Sevrin, On the BRST operator of $W$-string, preprint UG-2/93,
UCB-PTH-93/05 or LBL-33737 (March 1993) }

One point worth noticing is that contrary to the general belief the
nilpotent condition doesn't fix all the coefficients. What is not quite
interesting is the fact that these free parameters are all related the
freedom of canonically changing the ghost antighost fields. We effectively
proved a no-go theorem. As discussed in several
papers [\POPEB, \POPE, \LAS], for higher spin extended $W$-algebra, it
is often possible to split the  BRST operator into a sum
of mutually anticommuting and separately nilpotent operators.
If this is the case it is quite  easy to show that the BRST operator must
depend on some free parameters. If
$Q = Q_0 + \sum_{i=1}^n Q_i$  satisfy
$$ \{ Q_i, Q_j \} = 0, \qquad \hbox{ i, j} =0, 1, \cdots, n,
\eqn\wwd$$ the operator $\tilde{Q} = Q_0 + \sum_{i = 1}^n a_i Q_i$ is also
nilpotent and
depends on $n-1$ free parameters. Because our ansatz for the BRST charge is
the most general form and we found no free paramters in $Q$, this  proves
that in order to write $Q$ as the above form, one must use more inputs besides
the algebra, i.e. using an explicit realiztion of the algebra\footnote*{I
would like to thank Chris Pope to point out this to me.}.

\vglue .5cm

The author would like to thank Dr. K. Hornfeck, Prof. R. Iengo
and Dr. D. P. Li  for interesting discussions.
This work at SISSA/ISAS was supported by an INFN post-doctoral
fellowship.

\vglue .5cm

\vfill\eject
\refout
\bye